# Birefringence induced by antiferroelectric switching in transparent polycrystalline PbZr$_{0.95}$Ti$_{0.05}$O$_3$ film

Pranab Parimal Biswas,[1,2*] Cosme Milesi-Brault,[1,2,3,4] Alfredo Blázquez Martínez,[1,2,3] Naveen Aruchamy,[1,3] Longfei Song,[1,3] Veronika Kovacova,[2,3] Sebastjan Glinsek,[2,3] Torsten Granzow,[2,3] Emmanuel Defay,[2,3] Mael Guennou,[1,2,*]

[1]Department of Physics and Materials Science, University of Luxembourg, 41 rue du Brill, L-4422 Belvaux, Luxembourg.

[2]Inter-institutional Research Group Uni. lu–LIST on ferroic materials, 41 rue du Brill, L-4422 Belvaux, Luxembourg

[3]Materials Research and Technology Department, Luxembourg Institute of Science and Technology, 41 rue du Brill, L-4422 Belvaux, Luxembourg.

[4]Institute of Physics of the Czech Academy of Sciences, Na Slovance 1999/2, 182 21 Prague, Czech Republic

**Abstract:** The most characteristic functional property of antiferroelectric materials is the possibility to induce a phase transition from a non-polar to a polar phase by an electric field. Here, we investigate the effect of this field-induced phase transition on the birefringence change of PbZr$_{0.95}$Ti$_{0.05}$O$_3$. We use a transparent polycrystalline PbZr$_{0.95}$Ti$_{0.05}$O$_3$ film grown on PbTiO$_3$/HfO$_2$/SiO$_2$ with interdigitated electrodes to directly investigate changes in birefringence in a simple transmission geometry. In spite of the polycrystalline nature of the film and its moderate thickness, the field-induced transition produces a sizeable effect observable under a polarized microscope. The film in its polar phase is found to behave like a homogeneous birefringent medium. The time evolution of this field-induced birefringence provides information about irreversibilities in the antiferroelectric switching process and its slow dynamics. The change in birefringence has two main contributions, one that responds briskly (~ 0.5 s), and a slower one that rises and saturates over a period of as long as 30 minutes. Possible origins for this long saturation and relaxation times are discussed.

*Electronic mail: Authors to whom correspondence should be addressed: pranab.biswas@uni.lu, mael.guennou@uni.lu

Antiferroelectric (AFE) materials are related to the family of vastly used piezoelectric and ferroelectric (FE) materials. The concept of antiferroelectricity was first postulated by Kittel in 1951 and then experimentally demonstrated in PbZrO$_3$ (PZO) by Shirane *et al*. [1,2]. AFE materials as often described as being composed of two sublattices of dipoles with opposite directions, leading to a zero macroscopic polarization [3]. A sufficiently large electric field above a critical value ($E_{\text{AFE-FE}}$) can force a phase transition to a polar phase. This field-induced transition is the main defining feature of AFE that gives rise to their characteristic double hysteresis loop. For the last decade, a large portion of research has been dedicated to exploring this remarkable nature of AFEs for their potential applications in capacitors, actuators, and cooling devices [4–8].

In its bulk room-temperature phase, the archetypal AFE PZO is orthorhombic *Pbam* and undergoes a transition to a cubic $Pm\bar{3}m$ phase at 230 °C [3]. The structure of the field-induced polar phase of PZO has been under controversy. A number of phases have been proposed, yet, the rhombohedral phase seems to be the most accepted one [10–12]. Both the non-polar orthorhombic phase and polar rhombohedral phase of PZO are birefringent, more precisely optically biaxial and uniaxial, respectively. It is therefore natural to expect a change in birefringence at the phase transition, both in the orientation of the optical axes and the magnitude of birefringence. As a matter of fact, birefringence measurements have proven to be very insightful in general in studies of ferroic transitions with varying temperatures, including PZO [13]. The field-induced birefringence change has been extensively studied in oxide FE material, where they are known as electro-optic effect [14–17]. However, the birefringence change occurring along with antiferroelectric switching has barely been reported for oxide AFE materials. The development of transparent lead-based AFE and advancement in thin-film growth technologies provided electrical and optical characterization opportunities even in pure PZO, which otherwise would break down before the critical field is applied [18]. The relative lack of studies on this subject with the exception of Ref. [18,19], prompted us to delve into the field-induced AFE-FE phase transition via the birefringence effect. Keeping the potential application in transparent optoelectronic devices in mind, we utilize a transparent PbZr$_{0.95}$Ti$_{0.05}$O$_3$ (PZT) film to study the birefringence change caused by the field-induced AFE-FE phase transition in transmission geometry. The composition with 5 % of Ti has been selected in order to reduce the critical field.

A polycrystalline PZT film of thickness 1 µm was deposited by chemical solution deposition method on a fused-silica (SiO$_2$) substrate, coated with 23 nm of ALD deposited HfO$_2$ buffer layer. Before further processing, HfO$_2$ film was annealed at 700 °C and crystallized in the monoclinic phase [20]. To induce the preferred orientation, a seed layer of PbTiO$_3$ (PTO) was deposited on the substrate prior to PZT film deposition. The preparation of solutions for PZT and PTO followed a standard method [21]. The film was fabricated by spin coating the PZT solution on the PTO seed layer. Every deposited layer was dried for 3 min at 130 °C and pyrolyzed for 3 min at 350 °C. After every 4 layers of depositions, the film was crystallized in a rapid thermal annealing furnace at 700°C for 5 min in air. The process was repeated until the desired film thickness was reached. Platinum interdigitated electrodes (IDE) with a gap of 10 µm and a finger thickness of 5 µm were patterned with lift-off photolithography.

X-ray diffraction patterns of the PZT films (Bruker D8 Discover) are shown in Fig. 1 (a). The pattern reveals that the film crystallizes in the perovskite phase, without any impurity phases, and with pronounced orientation along <100> pseudo-cubic direction. The <100> pseudo-cubic direction could be associated with any of the three possible directions of the orthorhombic unit cell (<120>$_O$, <-120>$_O$, and <002>$_O$). However, upon closer inspection and comparing the XRD pattern with the standard powder diffraction as shown in Fig. 1 (a), we notice that the out-of-plane orientation of the film is specifically along <002>$_o$-direction (subscript pc and $o$ refer to pseudo-cubic and orthorhombic indexing, respectively) [22]. To estimate the (00$l$)$_{pc}$-orientation of the film, the Lotgering factor is calculated using the intensity relation of XRD peaks given in Ref. [23]. The Lotgering factor corresponding to (00$l$)$_{pc}$ orientation is approximately 0.86, indicating the film is well oriented. The in-plane orientation of the film however remains random.

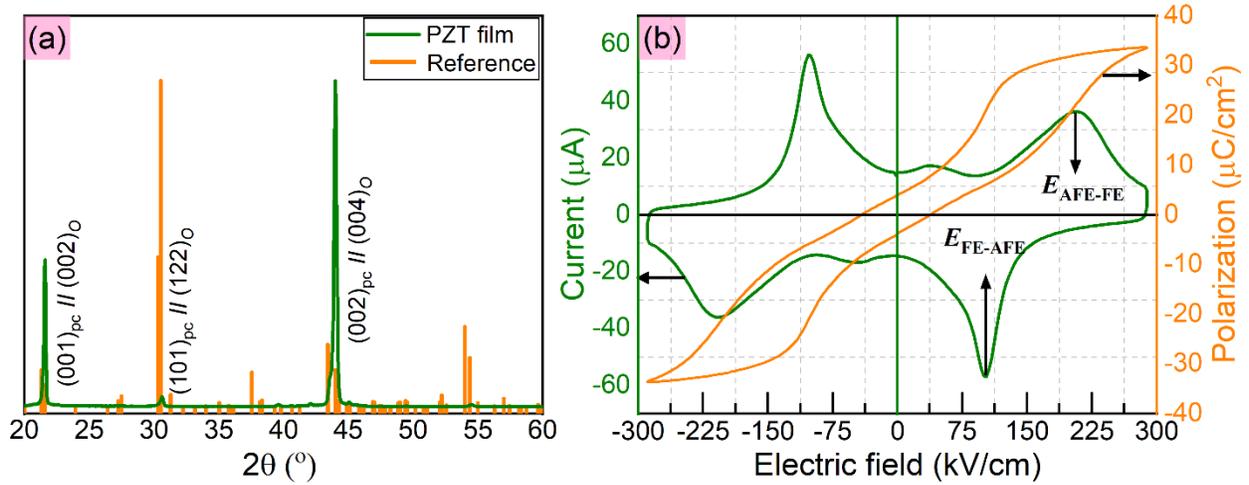

Fig. 1 (a) X-ray diffraction pattern, (b) Polarization ($P$) versus electric field ($E$) / Current ($I$) versus electric field ($E$) loops measured at 100 Hz.

The antiferroelectric hysteresis loop and the corresponding current curve were measured using a TF Analyzer 2000 from aixACCT by sending a bipolar triangular voltage ramp with a frequency of 100 Hz. The obtained result is shown in Fig. 1(b). The effective electric field across the IDE is derived from the applied voltage using the relation $E = V / (a + \Delta a)$ [24]. Here, $a$ is the gap between IDE fingers and $\Delta a = 1.324\ t_f$, with $t_f$ being film thickness. From the switching loops, a broad $E_{AFE\text{-}FE}$ was noted with a maximum transition field of ~ 205 kV/cm, and the back transition to the non-polar AFE phase ("back-switching") occurs at 94 kV/cm ($E_{FE\text{-}AFE}$). As discussed in Ref. [25] we do not expect a sharp current peak since the films contain a distribution of orientations in the plane of the film and therefore switching fields.

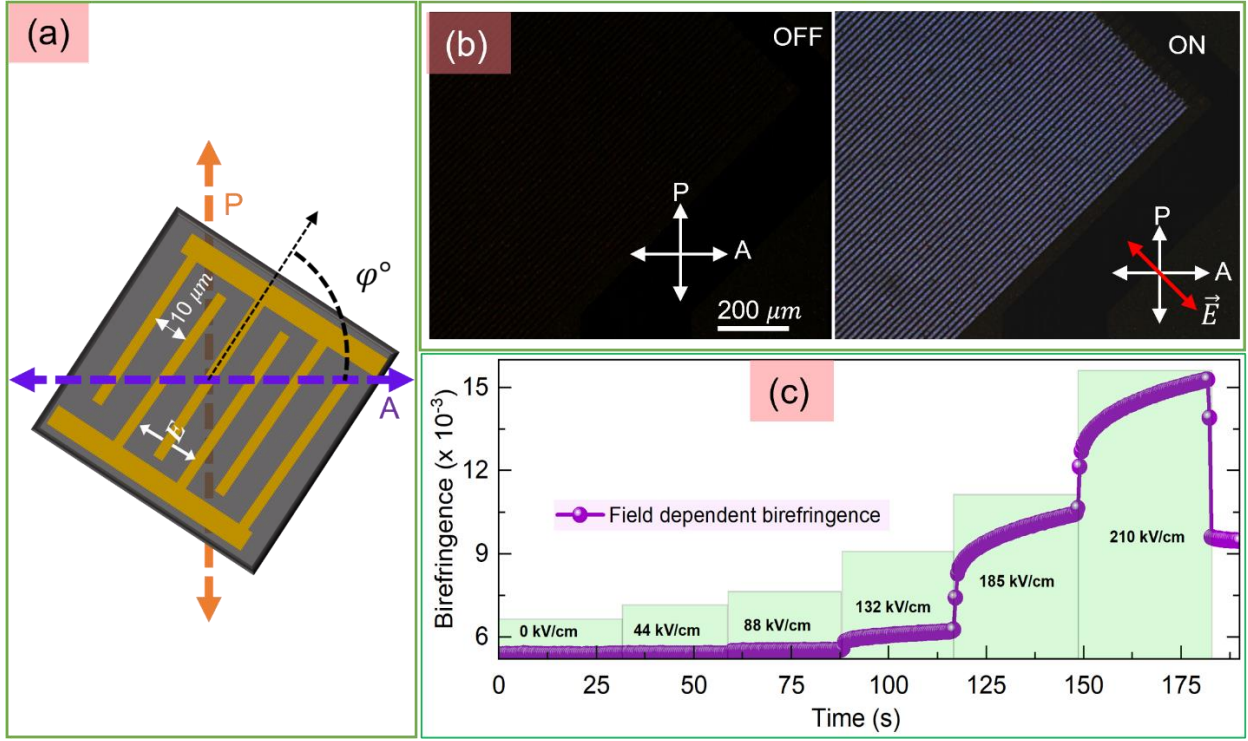

Fig. 2 (a) Schematic representation of birefringence measurement on PZT film under a polarizing microscope (top view), (b) Photographs of PZT film under the polarizing microscope in field-OFF and field-ON (205 kV/cm) states and (c) Electric-field dependence of birefringence. Here, A, P, and V ($E$) represent the light transmission axes of the analyzer, polarizer, and DC voltage (electric field), respectively.

For the birefringence measurement, a polarizing optical microscope was used (Leica DM2700M). Fig. 2 (a) demonstrates a schematic of a general arrangement of the experimental setup used for field-induced birefringence measurements. The light intensity transmitted through the sample was recorded at a rate of 100 frames per second and has been analyzed using image processing software [26]. An approximate area of 250 x 250 µm² was integrated in order to get an average intensity.

For a single crystal, the intensity transmitted through crossed polarizers [27] is given by:

$$I = I_o \sin^2(2\varphi) \sin^2(\delta/2) \qquad (1)$$

$$\delta = (2\pi/\lambda)\, \Delta n * t$$

Where, $I_o$ is the incident light intensity, $\varphi$ the angle between the polarizer and the axes of the optical indicatrix of the sample, $\delta$ the phase retardation, $\lambda$ the wavelength of the light, $\Delta n$ the crystal's birefringence, and $t$ is the sample thickness. When the permitted light polarization passing through the crystal is parallel with that of either the polarizer's or analyzer's axis, the crystal becomes dark. This is known as a position of extinction and is often used as a reference point. The maximum brightness/birefringence is observed when the optical axis of the crystal is at an angle of 45° with respect to both polarizer and analyzer. For our polycrystalline film with random in-plane orientation, the total intensity collected is the sum of the intensities from individual grains. The grain orientations are random in the plane but each grain is birefringent, so that a pristine films always transmits some light intensity. We therefore analyzed the transmitted intensity with the following formula;

$$I = I_{Background} + I_o \sin^2(2\varphi) \sin^2(\delta/2) \qquad (2)$$

In addition to the parameters above, here, $I_{Background}$ is the intensity measured outside the electrode region, and the angle $\varphi$ is defined as the angle between the fingers of the IDE and the Analyzer's axis, (Fig. 2A). We took 550 nm as the peak wavelength of our white light. $\Delta n$ becomes an effective birefringence that is a result of an orientational average.

Fig. 2 (b) displays images of PZT film under the polarizing microscope at $\varphi = 45°$, with and without field. A strong contrast between the electric field-OFF state and the electric field-ON state is apparent. To confirm that the observed birefringence is indeed linked to the phase transition, we performed this measurement with increasing values of the electric field. The results are displayed in Fig. 2 (c). Fields of 88 kV/cm and 132 kV/cm only caused a negligible response. The first prominent response is seen at 185 kV/cm which is much closer to the $E_{\text{AFE-FE}}$ phase transition field (205 kV/cm). The birefringence clearly increases with the field strength until 210 kV/cm. The field could not be increased beyond this value due to the electrical breakdown of the electrodes. The small response at the lower fields is probably because some of the crystallites start switching at lower fields as evident from the broad $E_{\text{AFE-FE}}$ peak in the $I(E)$ loop shown in Fig. 1 (b). However, the response is prominent only when the field is in the vicinity of or above the critical field. From the results discussed above, it is clear that the observed birefringence effect is indeed caused by the AFE-FE phase transition. The change in birefringence can therefore be seen as an effective electro-optic effect that sums up two main contributions: a change in intrinsic birefringence due to the change in crystal structure, but also a linear electro-optic effect that is forbidden in the AFE phase and becomes symmetry allowed in the polar phase.

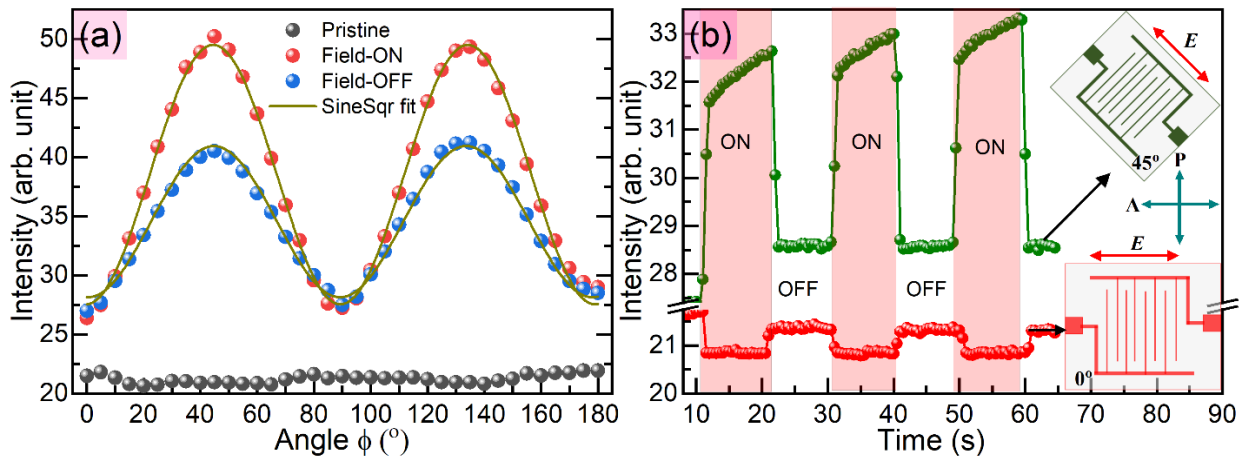

Fig. 3 (a) Azimuthal angular ($\varphi$) dependence of transmitted intensity for AFE phase (pristine), FE phase (field-ON @ 205 kV/cm), and remnant (field-OFF). (b) Field-induced (205 kV/cm) birefringence measurement at $\varphi = 0°$ and at $\varphi = 45°$. The time interval between each ON and OFF is 10s. The inset shows schematics of the top view of the sample positions for $\varphi = 45°$ and $\varphi = 0°$.

In order to better analyze the birefringence caused by field-induced AFE-FE phase transition, we first measured the transmitted light intensity as a function of the angle $\varphi$ with and without field (Fig. 3 (a)). For a pristine sample, the light intensity is constant and independent of the sample orientation. This behavior can be easily related to the polycrystalline nature of the film. The PZT film is highly oriented along the $[001]_{pc}$ // $[002]_O$-direction but otherwise isotropic in-plane. Orthorhombic PZO has principal refractive indices ($n_a < n_c < n_b$) along the directions of the crystallographic axes $a$, $b$, $c$ [28]. So, with light propagating along the $[002]_O$-direction under a polarizing microscope, the relevant birefringence for a single grain is $(n_b - n_a)$ [29]. By averaging over all possible grain orientations in the plane of the film, one obtains an optically isotropic behavior whereby some light intensity is transmitted in a way that is independent of the angle $\varphi$.

When the electric field corresponding to the AFE-FE phase transition (205 kV/cm) is applied, the transmitted intensity follows a characteristic sine-squared-shape dependence that is consistent with equation (2) and the behavior of a homogeneous birefringent medium. This also is the result of an average over different possible orientations of crystallites in the observed area: if we assume that the field-induced polar phase has a rhombohedral structure and that all grains are switched into the polar phase, then their optical axis lies along the $[111]_{pc}$ // $[042]_O$-direction that is closest to the field direction, which results in a net transmitted intensity. Remarkably, after the field is switched off, the light intensity no longer behaves like a pristine sample but still shows a significant magnitude and $\varphi$-dependence. This points to an irreversible process upon the first application of the field. In Fig. 3 (b), further measured birefringence are shown in two fixed positions ($\varphi = 0°$ and 45°) as a function of time while switching the DC-field ON and OFF in 10s steps. For $\varphi = 45°$ the intensity rises rapidly as soon as the field is turned on. The effect has a much smaller magnitude for the $\varphi = 0°$ position. In these conditions, the change in birefringence appears reproducible from one step to another after the initial switching.

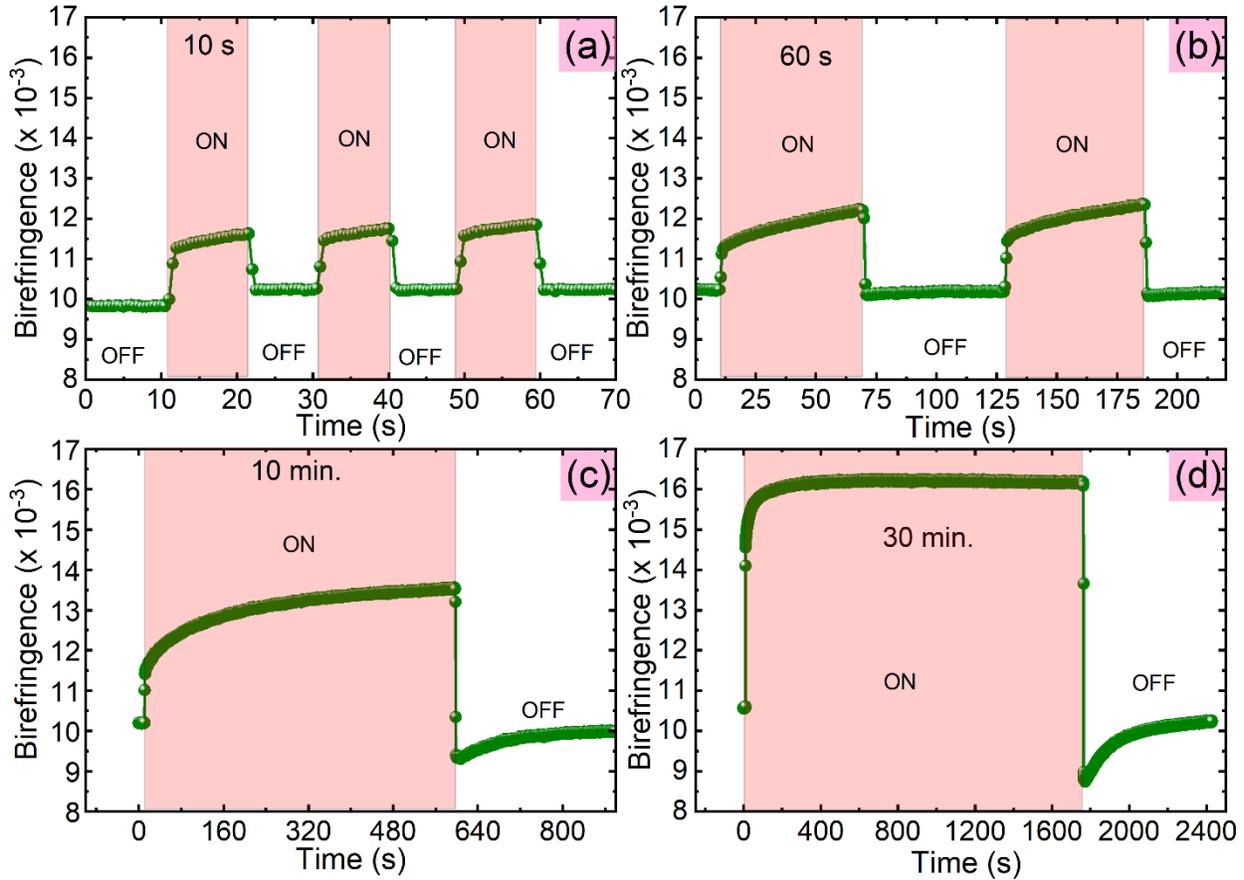

Fig. 4 Time-dependent birefringence measurement at $\varphi = 45°$ for various intervals between field-ON (205 kV/cm) and field-OFF states: (a)10 s, (b) 60 s, (c) 10 minutes, (d) 30 minutes.

Figure 3 (b) also shows that the birefringence during the field-ON duration is not constant over time but increases slowly for $\varphi = 45°$, with a trend towards saturation. To better understand it, we carried out time-dependent birefringence measurements for various time intervals between field-ON and field-OFF states as exhibited in Fig. 4. We note three important observations in these measurements: (i) an increase in birefringence that occurs in two steps when the field is switched on, with a first swift jump followed by a much slower increase, (ii) the presence of a remnant birefringence i.e., an irreversible contribution that remains even after the field is switched off, and

(iii) a significant undershoot upon back switching when the field is switched off, followed by a slow relaxation.

Switching of PZT-based antiferroelectric thin films under a DC electric field is reported to occur on a time scale of 6 ns to 10 ns, both forward (AFE-FE) and backward (FE-AFE) [30,31], i.e. it is instantaneous at the level of our measurement technique. The fast-responding component of the birefringence can therefore be associated with AFE-FE phase transition. The origin of the slow responding increase in birefringence could be linked to the growth and movement of newly formed domains and domain walls, respectively, in the polar phase under the DC electric field for a longer duration. To understand that, let us refer back to the assumption that the polar phase is rhombohedral. The transition of non-polar orthorhombic to the polar rhombohedral phase comes with the formation of many domain walls (DWs) because of the absence of a group-subgroup relation. In the rhombohedral phase, we would have 180°, 71°, and 109° DWs. Under the applied field, the ferroelectric domains are anticipated to grow until saturation. This argument is verified from frequency-dependent $P(E)$ / $I(E)$ hysteresis loop measurements where the saturation polarization ($P_s$) is seen to increase with a decrease in frequency (higher measurement time) without affecting the critical fields as displayed in Figure S1. The rate at which these DWs move could be affected by the presence of defects and dislocations through the pinning of DWs [32]. Moreover, 71° and 109° DWs being ferroelastic in nature, the growth is accompanied by the local stress and strain causing a delay in the saturation.

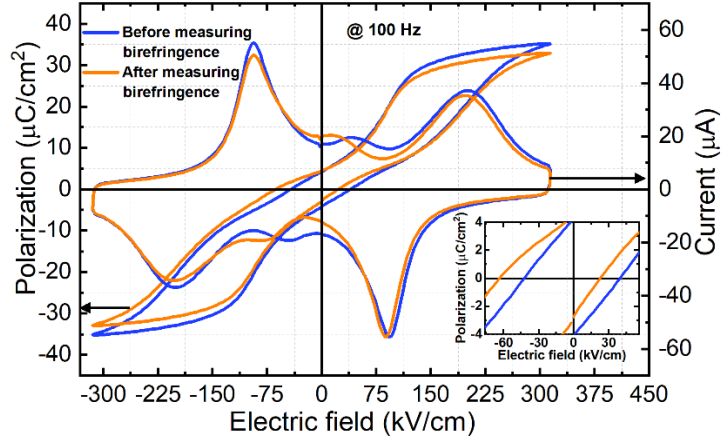

Fig. 5 $P(E) / I(E)$ hysteresis loops measured before (on a pristine film) and after the birefringent measurements. The inset shows a magnified view of the coercive field region.

The existence of a remnant birefringence points to an irreversible process, primarily, the irreversible FE domain formation due to the application of the high field. Fig. 5 shows $P(E) / I(E)$ hysteresis loops measured before and after the birefringence measurements. The decrease in the saturation polarization and the critical fields point toward the FE domain stabilization. This is further validated by bipolar fatigue measurements shown in Figure S2, where the effect is much more severe. From the evolution of $P(E) / I(E)$ hysteresis loops during the electric fatigue measurement, it is noticed that the $P_s$ and critical fields decrease while the remnant polarization ($P_r$) increases with an increase in the number of cycles. The double hysteresis loop nearly becomes a single loop after $10^5$ cycles, suggesting the stabilization of the FE domains upon bipolar field cycling [33]. Another potential source of remnant birefringence is the small but finite polarization value observed at zero electric fields in the $P(E)$ hysteresis loop. The non-zero polarization in the present PZT film could originate either from the presence of intrinsic weak ferroelectricity in PZO-based antiferroelectric films or the additional polar PTO seed layer in spite of its small thickness (13 nm) compared to the PZT film [34,35]. The presence of the FE domain in AFE PZO is not unprecedented [36]. In fact, recently, Aramberri et al. proposed that the ground state of PZO could well be ferri-electric [37]. In their first principle simulation based on density functional theory, the authors report that the most stable state of PZO is ferri-electric, among the known polymorph of this compound. Thus, one can't rule out the possibility that the finite remnant polarization observed at zero electric field in Fig. 1 (b) is due to the presence of this ferri-electric phase.

Finally, the reason for the undershoot at the back switching and the slow relaxation that follows is not clear at this stage. Hypotheses can be made based on the presence of a space charge effect, mobile charged defects like oxygen vacancies or defect dipoles that are known to form during the high temperature annealing of PZT film [38,39]. If we consider mobile charged defects, the application of a strong electric field for a long period could create a space charge effect, for example by migrating the charged defects near the grain boundaries, causing an internal bias field in the film. These defects could then re-distribute themselves after the field is switched off. Similarly, if we consider polar defects, the application of a strong electric field for a long period could align defect dipoles with the field that then do not switch back when the field is switched off but provides a restoring force towards a stable state that is partially polarized. The relaxation time in both cases may vary from tens of minutes to days [40–42]. A thorough analysis of the observed relaxation is beyond the scope of this paper, but we note that a similarly slow and isothermal transformation has been recently reported in a Nb-doped PZO crystal [43].

In summary, we studied the birefringence effect caused by field-induced AFE-FE phase transition in a transparent stack of antiferroelectric PbZr$_{0.95}$Ti$_{0.05}$O$_3$/PbTiO$_3$/HfO$_2$/SiO$_2$ thin film. We demonstrated that the observed birefringence is specifically induced by an AFE-FE phase transition. The azimuthal angle ($\varphi$) dependence of birefringence revealed that despite the polycrystalline nature, the PZT film in its field-induced polar phase behaves as a homogenous birefringent medium. The time dependence of birefringence was probed for various intervals between field-ON and field-OFF. It was found that there are three components in the observed birefringence effect corresponding to various factors: (i) a fast responding component associated with AFE-FE phase transition, (ii) a slow responding component due to ferroelectric and ferroelastic domains/domain walls growth/motion in the polar phase and (iii) a remnant birefringence owing to the finite remnant polarization in the film and field-induced permanent FE domains. More generally, this works shows that this relatively simple protocol for birefringence measurements can be used as a sensitive probe for field induced transitions in antiferroelectric thin films.

**Supplementary Materials**

See supplementary materials for more discussion on frequency dependent $P(E)$ / $I(E)$ hysteresis loops and fatigue measurements on PbZr$_{0.95}$Ti$_{0.05}$O$_3$ thin film.

**Acknowledgments**

This work is supported by the Luxembourg National Research Fund (INTER/ANR/16/11562984). The authors would like to thank Uros Prah for his help in commentary measurements.

**Data Availability**

The data that supports the findings of this study are available from the corresponding author upon reasonable request.

# References

[1]  C. Kittel, *Theory of Antiferroelectric Crystals*, Phys. Rev. **82**, 729 (1951).
[2]  G. Shirane, E. Sawaguchi, and Y. Takagi, *Dielectric Properties of Lead Zirconate*, Phys. Rev. **84**, 476 (1951).
[3]  E. Sawaguchi, H. Maniwa, and S. Hoshino, *Antiferroelectric Structure of Lead Zirconate*, Phys. Rev. **83**, 1078 (1951).
[4]  S. Ke, M. Ye, P. Lin, H. Huang, B. Peng, Q. Sun, F. Wang, X. Peng, and X. Zeng, *$PbZrO_3$-Based Antiferroelectric Thin Film Capacitors with High Energy Storage Density*, Int. J. Adv. Appl. Phys. Res. **1**, 35 (2014).
[5]  S.-T. Zhang, A. B. Kounga, W. Jo, C. Jamin, K. Seifert, T. Granzow, J. Rödel, and D. Damjanovic, *High-Strain Lead-Free Antiferroelectric Electrostrictors*, Adv. Mater. **21**, 4716 (2009).
[6]  X. Li, S.-G. (David) Lu, X. Chen, H. Gu, X. Qian, and Q. Zhang, *Pyroelectric and Electrocaloric Materials*, J Mater Chem C **1**, 23 (2012).
[7]  P. D. Thacher, *Electrocaloric Effects in Some Ferroelectric and Antiferroelectric Pb(Zr, Ti)$O_3$ Compounds*, J. Appl. Phys. **39**, 1996 (1968).
[8]  P. Vales Castro et al., *Origin of Large Negative Electrocaloric Effect in Antiferroelectric $PbZrO_3$*, Phys. Rev. B **103**, (2021).
[9]  M. Valant, *Electrocaloric Materials for Future Solid-State Refrigeration Technologies*, Prog. Mater. Sci. **57**, 980 (2012).
[10] B. Jaffe, W. R. Cook, and H. Jaffe, *CHAPTER 6 - PROPERTIES OF $PbTiO_3$, $PbZrO_3$, $PbSnO_3$, AND $PbHfO_3$ PLAIN AND MODIFIED*, in *Piezoelectric Ceramics* (Academic Press, 1971), pp. 115–134.
[11] S. E. Reyes-Lillo and K. M. Rabe, *Antiferroelectricity and Ferroelectricity in Epitaxially Strained $PbZrO_3$ from First Principles*, Phys. Rev. B **88**, 180102 (2013).
[12] J. Íñiguez, M. Stengel, S. Prosandeev, and L. Bellaiche, *First-Principles Study of the Multimode Antiferroelectric Transition in $PbZrO_3$*, Phys. Rev. B **90**, 220103 (2014).
[13] I. Lazar, A. Majchrowski, A. Soszyński, and K. Roleder, *Phase Transitions and Local Polarity above TC in a $PbZr_{0.87}Ti_{0.13}O_3$ Single Crystal*, Crystals **10**, 4 (2020).
[14] L. S. Kamzina and I. P. Raevskii, *Electric Field-Induced Birefringence in $Pb_{0.94}Ba_{0.06}Sc_{0.5}Nb_{0.5}O_3$ (PBSN-6) Solid-Solution Single Crystals*, Phys. Solid State **47**, 1143 (2005).
[15] W. Jo, H. Cho, T. W. Noh, B. I. Kim, D. Kim, Z. G. Khim, and S. Kwun, *Structural and Electro-optic Properties of Pulsed Laser Deposited $Bi_4Ti_3O_{12}$ Thin Films on MgO*, Appl. Phys. Lett. **63**, 2198 (1993).
[16] L. S. Kamzina, I. P. Raevskii, V. V. Eremkin, V. G. Smotrakov, and E. V. Sakhkar, *Evolution of the Optical Properties of $Pb_{0.94}Ba_{0.06}Sc_{0.5}Nb_{0.5}O_3$ (PBSN-6) Solid Solution Single Crystals Caused by a Static Electric Field*, Phys. Solid State **45**, 1112 (2003).
[17] S. Lee, T. W. Noh, and J. Lee, *Control of Epitaxial Growth of Pulsed Laser Deposited $LiNbO_3$ Films and Their Electro-optic Effects*, Appl. Phys. Lett. **68**, 472 (1996).
[18] F. Wang, K. K. Li, and G. H. Haertling, *Transverse Electro-Optic Effect of Antiferroelectric Lead Zirconate Thin Films*, Opt. Lett. **17**, 1122 (1992).
[19] F. Wang, K. K. Li, E. Furman, and G. H. Haertling, *Discrete Electro-Optic Response in Lead Zirconate Thin Films from a Field-Induced Phase Transition*, Opt. Lett. **18**, 1615 (1993).
[20] S. Glinsek et al., *Fully Transparent Friction-Modulation Haptic Device Based on Piezoelectric Thin Film*, Adv. Funct. Mater. **30**, 2003539 (2020).
[21] N. Godard, P. Grysan, E. Defay, and S. Glinšek, *Growth of {100}-Oriented Lead Zirconate Titanate Thin Films Mediated by a Safe Solvent*, J. Mater. Chem. C **9**, 281 (2021).
[22] H. Fujishita and S. Katano, *Re-Examination of the Antiferroelectric Structure of $PbZrO_3$*, J. Phys. Soc. Jpn. **66**, 3484 (1997).
[23] F. K. Lotgering, *Topotactical Reactions with Ferrimagnetic Oxides Having Hexagonal Crystal Structures—I*, J. Inorg. Nucl. Chem. **9**, 113 (1959).

[24] R. Nigon, T. M. Raeder, and P. Muralt, *Characterization Methodology for Lead Zirconate Titanate Thin Films with Interdigitated Electrode Structures*, J. Appl. Phys. **121**, 204101 (2017).
[25] C. Milesi-Brault, N. Godard, S. Girod, Y. Fleming, B. El Adib, N. Valle, S. Glinšek, E. Defay, and M. Guennou, *Critical Field Anisotropy in the Antiferroelectric Switching of PbZrO$_3$ Films*, Appl. Phys. Lett. **118**, 042901 (2021).
[26] R. W.S., *ImageJ*, (1997).
[27] M. Born and E. Wolf, *Chapter 14 - Optics of Crystals*, in *Principles of Optics: Electromagnetic Theory of Propagation Interference and Diffraction of Light*, 7th ed. (Cambridge University Press, 1999), p. 696.
[28] F. Jona, G. Shirane, and R. Pepinsky, *Optical Study of PbZr$_3$ and NaNbO$_3$ Single Crystals*, Phys. Rev. **97**, 1584 (1955).
[29] S. Huband, A. M. Glazer, K. Roleder, A. Majchrowski, and P. A. Thomas, *Crystallographic and Optical Study of PbHfO$_3$ Crystals*, J. Appl. Crystallogr. **50**, 378 (2017).
[30] T. K. Song, S. Aggarwal, Y. Gallais, B. Nagaraj, R. Ramesh, and J. T. Evans, *Activation Fields in Ferroelectric Thin Film Capacitors: Area Dependence*, Appl. Phys. Lett. **73**, 3366 (1998).
[31] S. S. N. Bharadwaja and S. B. Krupanidhi, *Backward Switching Phenomenon from Field Forced Ferroelectric to Antiferroelectric Phases in Antiferroelectric PbZrO$_3$ Thin Films*, J. Appl. Phys. **89**, 4541 (2001).
[32] C. Yang, E. Sun, B. Yang, and W. Cao, *Modeling Dynamic Rotation of Defect Dipoles and Poling Time Dependence of Piezoelectric Effect in Ferroelectrics*, Appl. Phys. Lett. **114**, 102902 (2019).
[33] W. Geng, Y. Liu, X. Lou, F. Zhang, Q. Liu, B. Dkhil, M. Zhang, X. Ren, H. He, and A. Jiang, *Polarization Fatigue in Antiferroelectric (Pb,La)(Zr,Ti)O$_3$ Thin Films: The Role of the Effective Strength of Driving Waveform*, Ceram. Int. **41**, S289 (2015).
[34] L. Pintilie, K. Boldyreva, M. Alexe, and D. Hesse, *Capacitance Tuning in Antiferroelectric–Ferroelectric PbZrO$_3$–Pb(Zr$_{0.8}$Ti$_{0.2}$)O$_3$ Epitaxial Multilayers*, New J. Phys. **10**, 013003 (2008).
[35] L. Pintilie, K. Boldyreva, M. Alexe, and D. Hesse, *Coexistence of Ferroelectricity and Antiferroelectricity in Epitaxial PbZrO$_3$ Films with Different Orientations*, J. Appl. Phys. **103**, 024101 (2008).
[36] X. Dai, J.-F. Li, and D. Viehland, *Weak Ferroelectricity in Antiferroelectric Lead Zirconate*, Phys. Rev. B **51**, 2651 (1995).
[37] H. Aramberri, C. Cazorla, M. Stengel, and J. Íñiguez, *On the Possibility That PbZrO$_3$ Not Be Antiferroelectric*, Npj Comput. Mater. **7**, 1 (2021).
[38] N. Mukhin, D. Chigirev, L. Bakhchova, and A. Tumarkin, *Microstructure and Properties of PZT Films with Different PbO Content—Ionic Mechanism of Built-In Fields Formation*, Materials **12**, 2926 (2019).
[39] M. Dawber and J. F. Scott, *A Model for Fatigue in Ferroelectric Perovskite Thin Films*, Appl. Phys. Lett. **76**, 1060 (2000).
[40] P. Erhart, P. Träskelin, and K. Albe, *Formation and Switching of Defect Dipoles in Acceptor-Doped Lead Titanate: A Kinetic Model Based on First-Principles Calculations*, Phys. Rev. B **88**, 024107 (2013).
[41] B. Akkopru-Akgun, W. Zhu, M. T. Lanagan, and S. Trolier-McKinstry, *The Effect of Imprint on Remanent Piezoelectric Properties and Ferroelectric Aging of PbZr$_{0.52}$Ti$_{0.48}$O$_3$ Thin Films*, J. Am. Ceram. Soc. **102**, 5328 (2019).
[42] C. Borderon, R. Renoud, M. Ragheb, and H. W. Gundel, *Dielectric Long Time Relaxation of Domains Walls in PbZrTiO$_3$ Thin Films*, Appl. Phys. Lett. **104**, 072902 (2014).
[43] D. Kajewski, I. Jankowska-Sumara, J.-H. Ko, J. W. Lee, S. F. U. H. Naqvi, R. Sitko, A. Majchrowski, and K. Roleder, *Long-Term Isothermal Phase Transformation in Lead Zirconate*, Mater. Basel Switz. **15**, 4077 (2022).